\documentclass[mathleft
]{an}
\usepackage{graphicx}
\usepackage{times}
\begin{document}

\Pagespan{789}{}
\Yearpublication{2006}%
\Yearsubmission{2005}%
\Month{11}%
\Volume{999}%
\Issue{88}%

\title{ULX behaviour: the ultraluminous state, winds \& interesting anomalies}

\author{T.\,P. Roberts\inst{1}\fnmsep\thanks{Corresponding author:
  \email{t.p.roberts@durham.ac.uk}\newline}
\and  M.\,J. Middleton\inst{2} \and A.\,D. Sutton\inst{3} \and M. Mezcua\inst{4} \and D.\,J. Walton\inst{5,6} \and L.\,M. Heil\inst{7}
}
\titlerunning{ULX behaviour}
\authorrunning{T.\,P. Roberts et al.}
\institute{
Centre for Extragalactic Astronomy, Department of Physics, Durham University, South Road, Durham DH1 3LE, UK
\and 
Institute of Astronomy, Madingley Road, Cambridge CB3 0HA, UK
\and 
Astrophysics Office, NASA Marshall Space Flight Center, ZP12, Huntsville, AL 35812, USA
\and
Harvard-Smithsonian Centre for Astrophysics, 60 Garden Street, Cambridge, MA 02138, USA
\and
NASA Jet Propulsion Laboratory, 4800 Oak Grove Drive, Pasadena, CA 91109, USA
\and 
California Institute of Technology, 1200 East California Boulevard, Pasadena, CA 91125, USA
\and
Astronomical Institute Anton Pannekoek, Science Park 904, 1098 XH, Amsterdam, Netherlands}

\received{9 Sep 2015}
\accepted{}
\publonline{later}


\abstract{Recent evidence - in particular the hard X-ray spectra obtained by {\it NuSTAR\/}, and the large amplitude hard X-ray variability observed when ULXs show soft spectra - reveals that common ultraluminous X-ray source (ULX) behaviour is inconsistent with known sub-Eddington accretion modes, as would be expected for an intermediate-mass black hole (IMBH).  Instead, it appears that the majority of ULXs are powered by super-Eddington accretion onto stellar-mass black holes. Here, we will review work that delves deeper into ULX spectral-timing behaviour, demonstrating it remains consistent with the expectations of super-Eddington accretion.  One critical missing piece from this picture is the direct detection of the massive, radiatively-driven winds expected from ULXs as atomic emission/absorption line features in ULX spectra; we will show it is very likely these have already been detected as residuals in the soft X-ray spectra of ULXs.  Finally, we will discuss ULXs that do not appear to conform to the emerging ULX behaviour patterns.  In particular we discuss the implications of the identification of a good IMBH candidate as a background QSO; and the confirmation of an IMBH/ULX candidate in the galaxy NGC 2276 via the radio/X-ray fundamental plane.}

\maketitle

\section{Introduction}
After a decade and a half of detailed study, primarily from space-based X-ray observatories but with input from a plethora of multi-wavelength facilities, the nature of ultraluminous X-ray sources (ULXs) is beginning to be revealed.\footnote{For the most recent review of ULXs see Feng \& Soria (2011).}  A debate has long raged over whether ULXs harbour a new class of intermediate-mass black hole (IMBH), with masses in the range $\sim 10^2 - 10^5 ~M_{\odot}$, that accrete in a familiar sub-Eddington mode, or whether ULXs instead represent a population of super-Eddington accretors, in which case we are most likely observing stellar-mass black holes (masses $\sim 5 - 20 ~M_{\odot}$)\footnote{The upper limit on this mass range might extend as high as $\sim 100 ~M_{\odot}$ in certain scenarios; see e.g. Belczynski et al. (2010).} in a new and very extreme mode of accretion.  The first masses inferred dynamically from optical data reveal two different ULXs to be powered by compact objects with masses in the stellar-mass black hole range (Liu et al. 2013; Motch et al. 2014).  A transient ULX in M31 has also been constrained to the same mass range (Middleton et al. 2013).  However, it is also becoming clear that ULXs are a heterogeneous population, with the detection of pulsations from M82 X-2 demonstrating that at least this ULX is powered by accretion onto a neutron star (Bachetti et al. 2014).  It also appears possible that some residual population of IMBHs could be hidden amongst the wider ULX population, with the best current candidates being the most luminous ULXs (e.g. Farrell et al. 2009, Sutton et al. 2012).

\begin{figure}
\centering
\includegraphics[width=65mm]{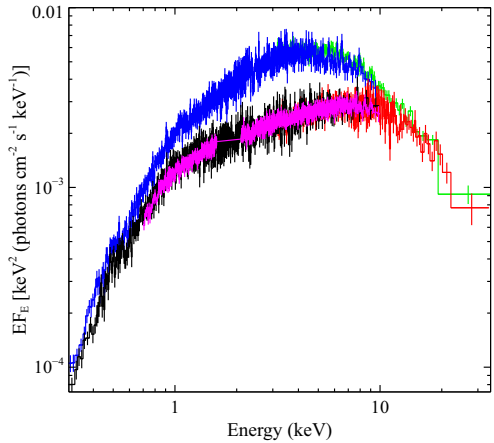}
\caption{Multi-mission spectroscopy of the ULX Ho IX X-1 in two separate epochs, unfolded from the instrument responses.  The lower flux data are from {\it XMM-Newton\/} (black), {\it Suzaku\/} (magenta) and {\it NuSTAR\/} (red), with the higher flux data from {\it XMM-Newton\/} (blue) and {\it NuSTAR\/} (green).  The spectral turnover is clear in both epochs.  Figure from Walton et al. (2014).}
\label{nustarspec}
\end{figure}

Despite the likely heterogeneity of the overall ULX population, it is becoming clearer that many of the best ULX datasets appear to show some commonalities in their properties that emphasise their difference from other accreting black holes (BHs).  Foremost amongst these is a near-ubiquitous downturn in their spectra at energies of a few keV, first detected in {\it XMM-Newton\/} data (e.g. Roberts et al. 2005, Stobbart, Roberts \& Wilms 2006) and latterly more spectacularly demonstrated by the harder bandpass of {\it NuSTAR\/} data (Fig.~\ref{nustarspec}; e.g. Bachetti et al. 2013, Walton et al. 2014).  Most high quality ULX spectra can be described by a combination of a hard component showing this downturn and a soft excess; such a spectrum is inconsistent with any of the classic BH accretion states.  Crucially for our understanding of these properties, Motch et al. (2014) confirm that they are displayed by a stellar-mass black hole accreting in a super-Eddington mode.  In this paper we will briefly review these and other properties, that demarcate the putative new ultraluminous accretion state, and discuss our recent results that provide further insight into the physics of this super-Eddington accretion mode.  We will also mention examples where the detection of ULXs with properties that differ from the emerging patterns of ULX behaviour may be highlighting ULXs with a different underlying nature.

\section{The ultraluminous state}
Early {\it XMM-Newton\/} spectra of ULXs determined the presence of a soft excess in many ULX spectra; as the hard component appeared power-law-like, the cool ($\sim 0.2$ keV) disc-like soft excess was initially interpreted as an accretion disc around an IMBH (e.g. Miller et al. 2003).  However, subsequent deeper ULX observations showed that the hard component appears to turn over at energies of a few keV in sufficiently high signal-to-noise data, producing spectra that can be well represented by two thermal components, not what would be expected from an IMBH accreting below the Eddington limit (Stobbart et al. 2006).  Instead, it was suggested that this unusual spectral morphology might represent a new, {\it ultraluminous accretion state\/} for BHs experiencing super-Eddington accretion (Roberts 2007).  Further inspection of the high signal-to-noise ULX spectra by Gladstone, Roberts \& Done (2009) revealed three distinct spectral shapes (see their Figure 9), subsequently called the {\it broadened disc\/} (hereafter BD), {\it hard ultraluminous\/} (HUL) and {\it soft ultraluminous\/} (SUL) regimes by Sutton, Roberts \& Middleton (2013b).  Clearly, if we are to understand the ULX phenomenon, then understanding both the behaviour in, and evolution between, these regimes could hold the key.

\subsection{Spectral regimes and coarse variability properties}

\begin{figure}
\centering
\includegraphics[width=65mm]{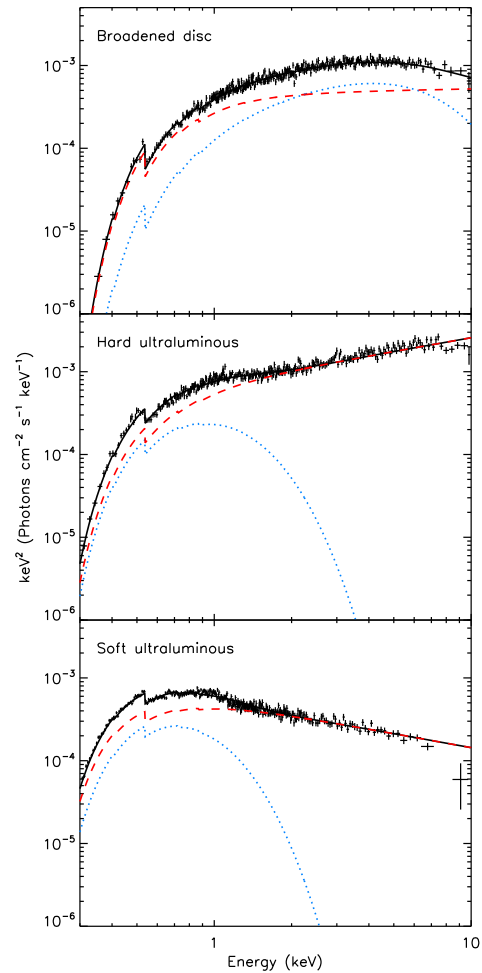}
\caption{Example spectra for the three ultraluminous state regimes.  {\it XMM-Newton} PN data is shown in black, rebinned to 10$\sigma$ significance for clarity.  The data are fitted with a simple multicolour disc blackbody plus power-law continuum model (shown separately as blue dotted and red dashed lines respectively), the parameterisation of which is used to classify the different regimes.  Figure from Sutton et al. (2013b).}
\label{ulxregimes}
\end{figure}  

\begin{figure*}
\includegraphics[width=56mm]{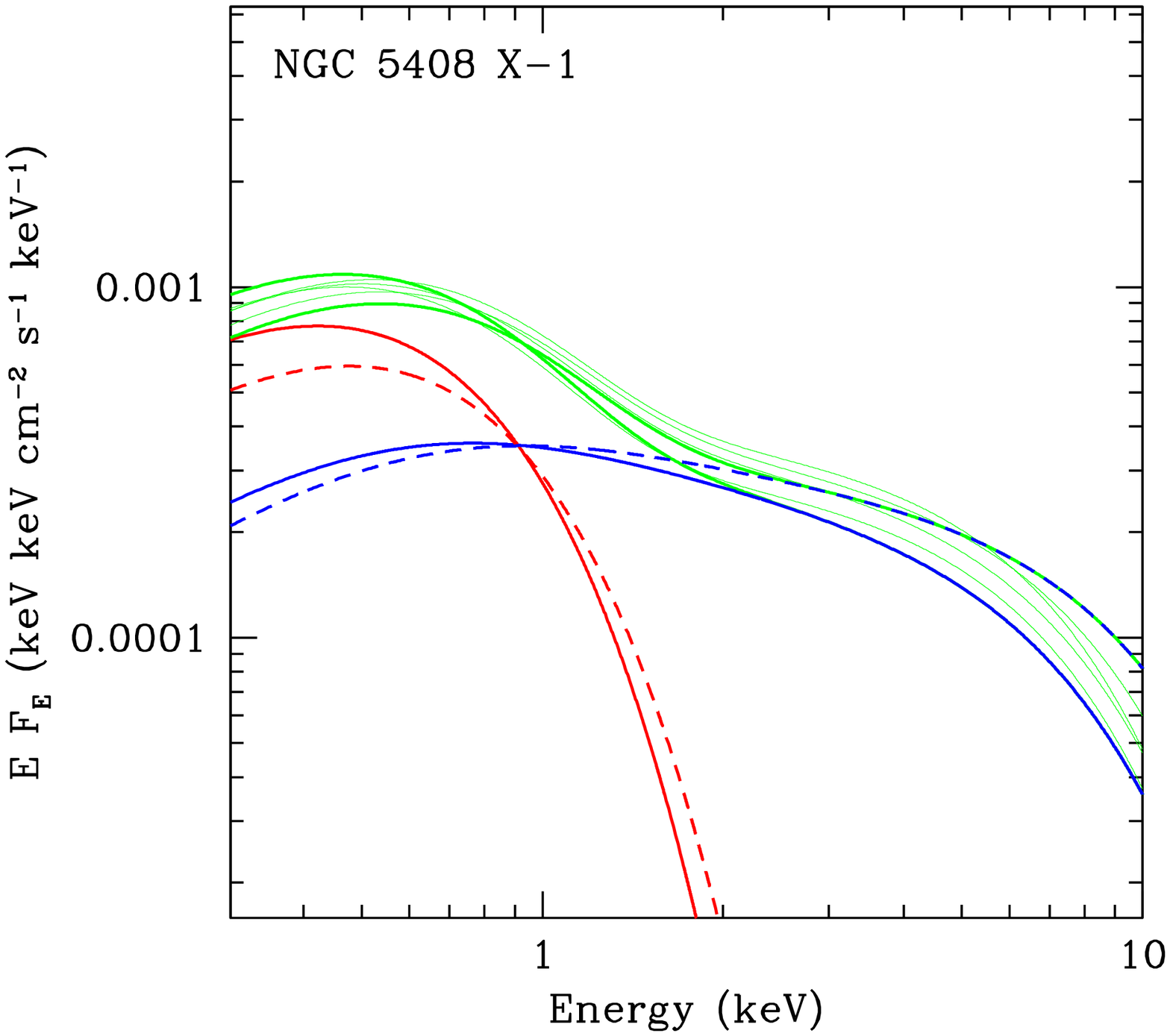}
\includegraphics[width=56mm]{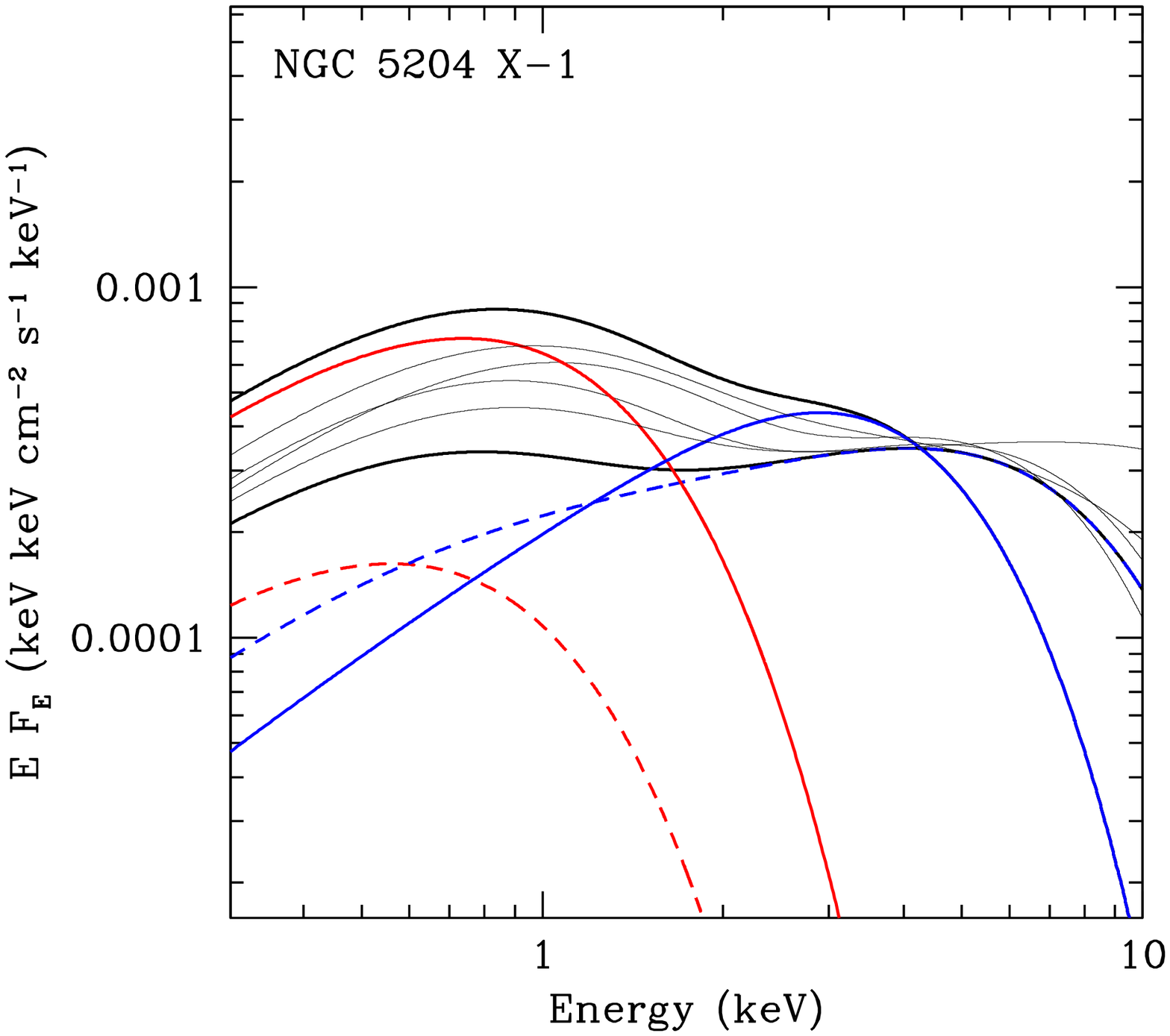}
\includegraphics[width=56mm]{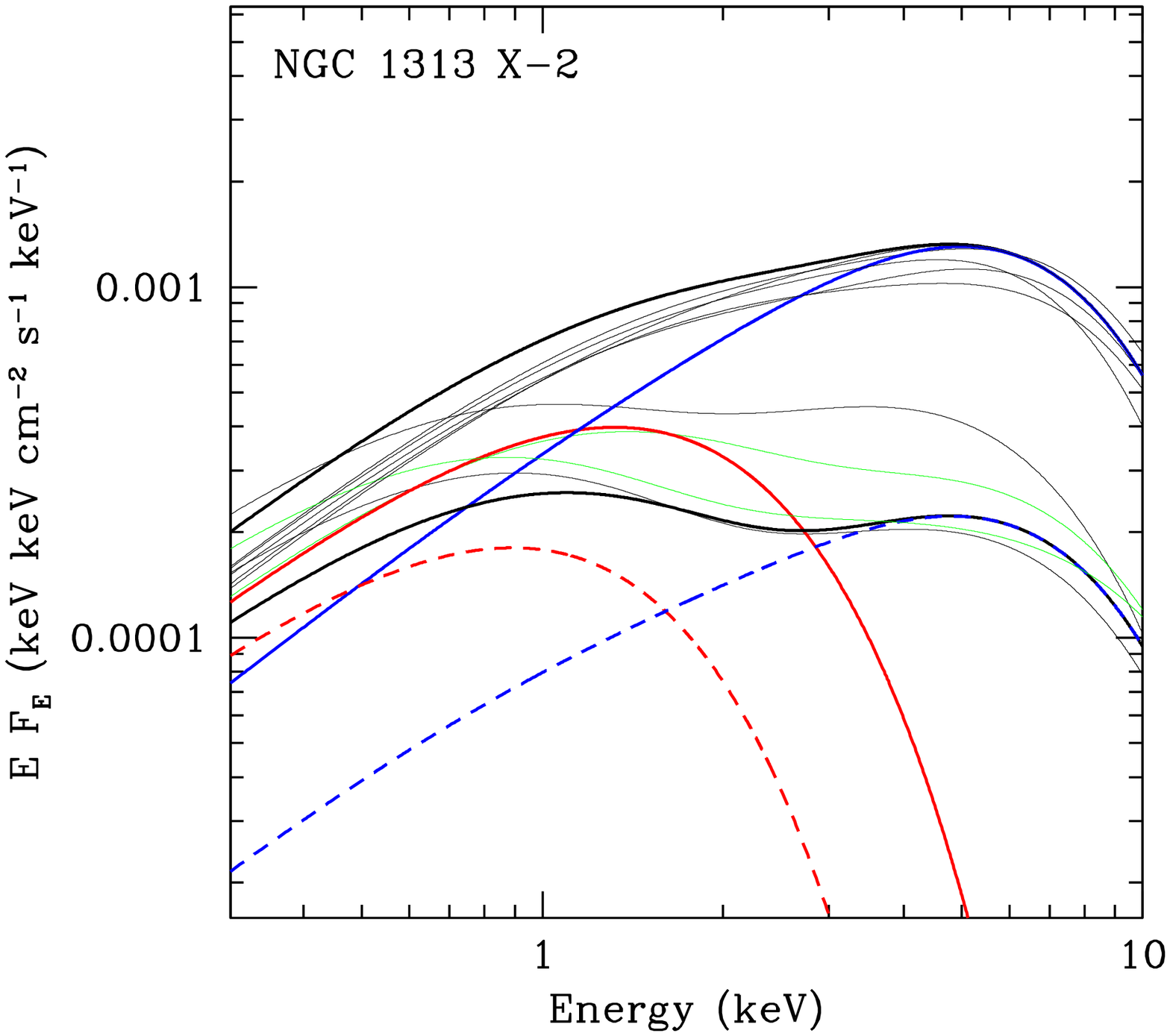}
\caption{The spectral variability of three ULXs.  Here we show the best fitting, de-absorbed continuum models obtained from multiple epochs of {\it XMM-Newton\/} data.  Where significant flux variability within an observation is observed we plot the model in green, otherwise it is plotted in black.  We also plot the underlying multicolour disc blackbody (DISKBB in {\sc xspec}) and Comptonisation (NTHCOMP) components of the models in red and blue respectively, for the most luminous (solid lines) and least luminous (dashed lines) observations of each ULX.  Figure from Middleton et al. (2015a).}
\label{ulxspecvar}
\end{figure*}

Example spectra showing the differences between the three regimes are shown in Fig.~\ref{ulxregimes}.  In Sutton et al. (2013b) we developed a simple method of distinguishing them based on a multi-colour disc plus power-law fit to {\it XMM-Newton\/} EPIC data (see figure 2 of that paper), and used this to investigate their simple properties in terms of flux, hardness and fractional rms variability (figures 3, 5 \& 6 of Sutton et al. 2013b).  In short, we found that BD objects dominate the population below a luminosity of $3 \times 10^{39} \rm ~erg~s^{-1}$, consistent with a population of stellar-mass BHs at mildly super-Eddington rates.  There are some BD spectra above this threshold, but it is unclear whether they constitute {\it bona-fide\/} BD objects -- implying larger stellar BHs -- or HUL objects with very weak soft excesses.  Most objects above $3 \times 10^{39} ~\rm erg~s^{-1}$ are in either the HUL or SUL states, and crucially they co-exist at similar luminosities.  All ULXs in the HUL regime, and almost all in the BD regime, show low fractional rms ($F_{\rm var} < 10\%$); however, some SUL regime objects show a high fractional rms ($F_{\rm var} \approx 10 - 30\%$), with this variability predominantly seen above 1 keV, although it is not persistently seen in all observations of all objects.

\subsection{Super-Eddington models and spectral evolution}

In the remainder of this work we will focus specifically on the HUL and SUL regimes.  If the BD regime is mildly super-Eddington, it seems likely that the HUL/SUL regimes are when stellar-mass BHs become strongly super-Eddington (supercritical).  In such a regime we would expect to see two dominant effects: the disc itself should become geometrically thick as its interior becomes advection dominated, and a very massive radiatively-driven wind should be blown from its surface.  This should produce a funnel-like structure in the central regions of the ULX, where a low density region along the BH rotational axis is bounded by the inner disc and optically-thick wind material (e.g. Poutanen et al. 2007; Kawashima et al. 2012).  The two-component X-ray spectrum is then naturally explained by the harder component originating in emission from the inner disc, and the softer component being  thermal emission from the optically-thick wind.\footnote{There is still some debate about this conclusion, particularly for the soft component, see Miller et al. (2014)}  Poutanen et al. (2007) noted that the appearance of a ULX will depend on the angle it is viewed at in this model; we suggested in Sutton et al. (2013b) that this explains the difference between SUL and HUL regimes, with the HUL viewed at low inclination (i.e. close to face-on) and so dominated by the inner-disc, while the SUL is viewed at higher inclination where wind emission dominates.  The similar luminosities observed from both regimes supports the notion that inclination rather than accretion rate is the key distinction between the two regimes.

In Middleton et al. (2015a) we revisited this model and described its implications in greater detail.  In particular, we discussed how changes in accretion rate at different viewing angles should affect the observed spectra.  Each component will behave differently depending on viewing angle; the wind is a roughly isotropic emitter, so the soft emission (to first order) depends on the accretion rate.  However, the hard component should scatter off the wind, meaning that if the wind increases then the expected narrowing of the funnel (cf. King 2009) should result in increased geometrical beaming of this component, so the hard emission will increase faster than the soft for a viewpoint in the beam; but outside the beam the observed spectrum should become softer, as the wind increases, but hard emission is scattered away from the line of sight.  These effects are precisely what we do see from observed ULXs with multi-epoch {\it XMM-Newton\/} spectra (Middleton et al. 2015a).  In particular, in Fig.~\ref{ulxspecvar} we show 3 objects: NGC 5408 X-1 appears to be seen at higher inclination, with a dominant wind but little long-term spectral evolution; NGC 5204 X-1 appears to be at moderate inclination, seen at first as HUL, but as its luminosity increases it becomes increasingly softer, transiting to SUL, consistent with the funnel narrowing at higher accretion rates such that the wind crosses the line-of-sight; finally NGC 1313 X-2 starts as HUL and becomes increasingly harder, consistent with a very low inclination angle.

\subsection{Variability induced by a clumpy wind?}
\begin{figure}
\centering
\includegraphics[width=65mm]{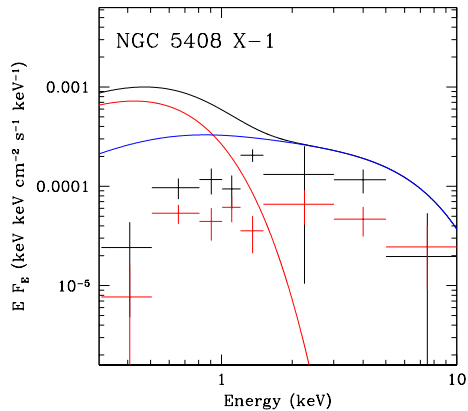}
\caption{Covariance spectra of NGC 5408 X-1 relative to the 1.5-3 keV band, over two timescales: long (red, 0.9-3 mHz) and short (black, 3-200 mHz).  These are plotted against the best-fitting, de-absorbed, time-averaged continuum model as per Fig.~\ref{ulxspecvar}.  Figure from Middleton et al. (2015a).}
\label{5408cv}
\end{figure}

The super-Eddington models also provide a very natural explanation for the high levels of variability in the SUL regime, and not the others: this is an extrinsic effect, imprinted on the emission from the inner-disc by an inhomogeneous wind in which clumps of material cross the line-of-sight, scattering away hard photons (Middleton et al. 2011; Takeuchi, Ohsuga \& Mineshige 2013).  In Middleton et al. (2015a) we investigated this possibility by extracting covariance spectra for several bright ULXs; in each case the spectrum of the varying emission matched closely to that of the hard spectral component (see example in Fig.~\ref{5408cv}).  This demonstrates that the variation is limited solely to the hard component, as would be expected from obscuration and scattering by a clumpy wind.

\section{Winds}
The above interpretation is strongly predicated upon the presence of a massive, outflowing wind, driven by the intense radiation pressure of the supercritical accretion disc.  But do we have any direct evidence for the presence of a wind?  The best direct evidence would be in the form of emission and/or absorption features from different atomic species in the wind.  An obvious place to look for these is the Fe K band, given the isolation of these features and their strength in stellar BHs and AGN; but none were detected in deep {\it Suzaku\/} observations of Ho IX X-1, chosen as a target because it offers the best signal-to-noise of a spatially isolated ULX at 6 keV and its intrinsically hard spectrum offers the best opportunity of observing reflection features (Walton et al. 2013).  However, in the context of the model described above this should perhaps be no surprise; Ho IX X-1 is a HUL object, and so we view it down its funnel where we do not expect to observe through a significant wind.

The objects that we do view through their wind are the SUL objects and, rather encouragingly, it has long been known that soft ULXs show residuals to their continuum spectral fits (e.g. Stobbart et al. 2006).  In Middleton et al. (2014) we investigated the residuals in the archetypal SUL objects NGC 5408 X-1 and NGC 6946 X-1, and found that they can be explained as broad absorption lines from a partially ionised and blueshifted ($v \approx 0.1 c$) medium -- consistent with absorption from the optically thin phase of a turbulent outflowing wind, located outside the last scattering surface of the optically-thick region.

\begin{figure}
\centering
\includegraphics[width=80mm]{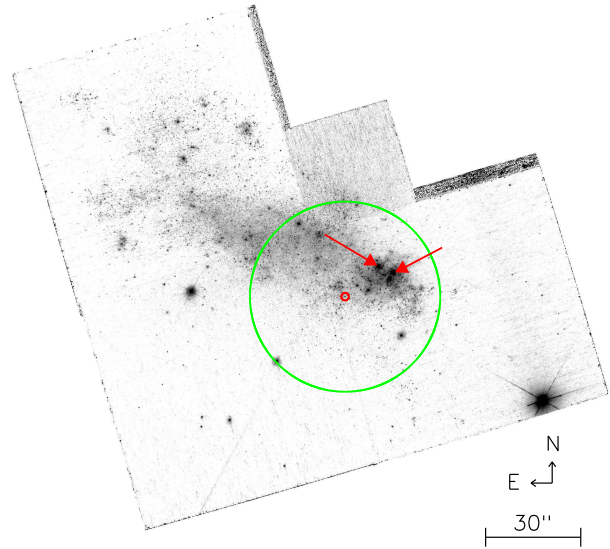}
\caption{{\it HST\/} WFPC2 F336W image of NGC 5408.  The footprint of a typical source data extraction region for {\it XMM-Newton\/} EPIC data is shown as the green circle, with that of {\it Chandra\/} ACIS data shown in red.  The main star forming regions in NGC 5408 are indicated by the red arrows.  Figure from Sutton et al., submitted.}
\label{5408img}
\end{figure}

However, this is not a unique solution to fitting the residuals, as they may also be fitted by the emission from an optically-thin thermal plasma.  This has led to the residuals regularly being dismissed as unrelated to ULXs in previous studies, where they are assumed to be due to hot gas in the host galaxy, perhaps related to its star formation (e.g. Strohmayer et al. 2007; Miller et al. 2013).  In new work (Sutton, Roberts \& Middleton, submitted), we show that {\it Chandra\/} is easily able to spatially resolve the position of the ULX from the main star formation regions in NGC 5408 that lie within the {\it XMM-Newton\/} footprint for the ULX, as shown in Fig.~\ref{5408img}.  Yet, in excess of 65\% of the residual emission (as modelled by a thermal plasma) remains spatially unresolved by {\it Chandra\/}, and so is very likely located in the proximity of the ULX.  We can therefore rule out star-formation related plasma for this ULX, and instead note the residuals are probably related to NGC 5408 X-1 itself.

\begin{figure}
\centering
\includegraphics[width=65mm]{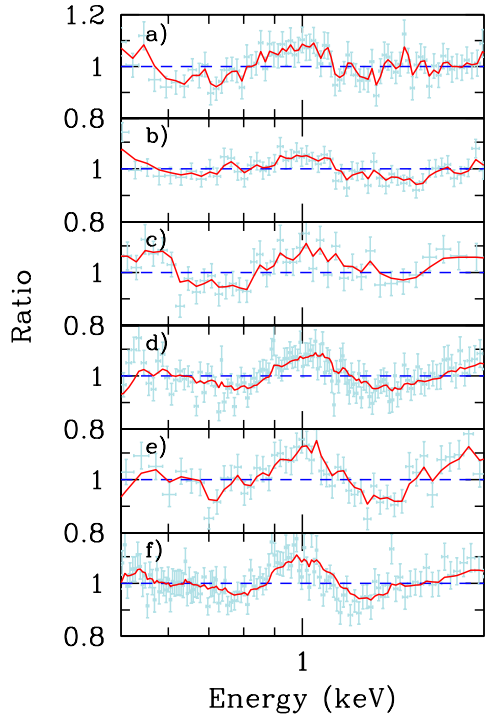}
\caption{Soft X-ray residuals to continuum models for 6 ULXs.  The objects are: (a) NGC 1313 X-1; (b) Ho IX X-1; (c) Ho II X-1; (d) NGC 55 ULX-1; (e) NGC 6946 X-1; and (f) NGC 5408 X-1.  The spectral data is shown in light blue and plotted as a ratio to the best fitting model, and an exponentially smoothed function is plotted in red to highlight the commonality in the shape of the residuals.  Figure from Middleton et al. (2015b).}
\label{resids}
\end{figure}

We have further investigated whether these residuals are related to the ULXs in a second new work (Middleton et al. 2015b).  In this we demonstrate that the residuals appear to have a very similar form across six different ULXs with high signal-to-noise soft X-ray spectra, which we show here as Fig.~\ref{resids}.  This argues that they have a common origin, related to the ULXs.  We further investigated this origin by examining how the residuals change with spectral hardness in NGC 1313 X-1, which has the largest dynamical range in its hardness of any of the ULXs with detectable residuals.  We find a clear anti-correlation between hardness and the depth of the features.  This rules out an origin in the reflection of a hard primary continuum, where we would expect stronger features if the spectrum becomes harder.  However it is consistent with wind models, where for moderate-to-high inclination angles an increase in accretion rate leads to a stronger wind, which both softens the spectrum and deposits more material into the optically thin medium surrounding the ULX, that deepens the absorption features.

\section{Interesting Anomalies}

The distinct spectral and temporal characteristics of the ULX regimes provides us with a sufficient depth of behaviours to begin to test models of super-Eddington accretion.  On a purely observation level, they also provide a template for the known behaviour of ULXs that are powered by stellar-mass BHs at super-Eddington rates.  It is therefore highly interesting to search for ULXs that do not appear to behave in the same way, as these may be powered by different underlying objects.  The primary example of this is the most luminous known ULX, ESO 243-49 HLX-1, whose regular outbursting behaviour appears to show it transition between classic sub-Eddington low/hard and high/soft modes (e.g. Servillat et al. 2011); a combination of its extreme peak luminosity and these apparently sub-Eddington states mean it is an excellent candidate for hosting an IMBH.\footnote{Although the outbursting behaviour of ESO 243-49 HLX-1 may also suggest other interpretations of its nature, see Lasota, King \& Dubus 2015.}  We discuss the most luminous ULXs as a class below, but note that an extreme luminosity is not a prerequisite for finding ULXs with unusual behaviour; we have recently found an object in M51 with fairly ordinary ULX luminosities but that displays a very hard spectrum and high fractional rms variability, reminiscent of a classic hard state, implying this object is a candidate IMBH (Earnshaw, these proceedings).

In Sutton et al. (2012) we examined a small sample of extreme ULXs, with peak luminosities above $5 \times 10^{40} \rm ~erg~s^{-1}$, and concluded that their observational properties -- in particular their hard power-law-like spectra and fractional rms variabilities of $\sim 10 - 20\%$ -- appeared consistent with a population of IMBHs in a classic sub-Eddington hard spectral state.  However, the data in most cases was only moderate quality, so there was some level of doubt over this interpretation.  Indeed, subsequent studies of the one object in the sample with relatively high signal-to-noise data have revealed a subtle spectral turnover, consistent with an object in the ultraluminous state (Sutton et al. 2013a; Walton et al. 2015).  It is therefore unclear whether the other objects in the sample would remain IMBH candidates if we were able to obtain high enough quality data from them.  Indeed, some may be background interlopers, as we have since discovered by studying the optical counterpart of the most luminous IMBH candidate in the extreme ULX sample, IC 4320 HLX, that turned out to be a $z \sim 2.8$ QSO behind the dusty lenticular galaxy IC 4320 (Sutton et al. 2015).  We also noted that, other than ESO 243-49 HLX-1, most ULXs that reached the hyperluminous regime had peak luminosities not much above $10^{41} \rm ~erg~s^{-1}$, consistent with a combination of the maximal radiative output from super-Eddington accretion and the largest possible stellar-mass black holes (Ohsuga \& Mineshige 2011; Zampieri \& Roberts 2009), so not all HLXs need contain IMBHs.

\begin{figure}
\centering
\includegraphics[width=75mm]{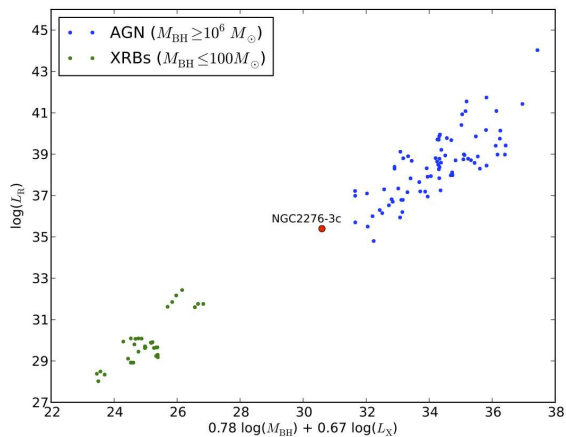}
\caption{The fundamental plane of black hole accretion from G{\"u}ltekin et al. (2009), with stellar-mass BHs (green dots) and AGN (blue dots) taken from Merloni et al. (2003).  A clear gap is observed in the IMBH range, where NGC 2276-3c is located.  Figure from Mezcua et al. (2015).}
\label{XRFP}
\end{figure}

However, one of the extreme ULXs has subsequently been shown to be an excellent IMBH candidate.  If these objects are IMBHs in a hard state they should be radio-bright, due to the presence of a persistent jet, so we undertook a radio study of these objects using the VLA to look for this emission (Mezcua et al. 2013).  We found that the ULX NGC 2276 3c was surrounded by a large radio nebula, and subsequent VLBI observations revealed a compact radio core to this nebula (Mezcua et al. 2015).  The acquisition of simultaneous {\it Chandra\/} observations to the VLBI data enabled us to place this extreme ULX on the X-ray/radio fundamental plane, and so determine it to have a mass of $\sim 5 \times 10^4 ~M_{\odot}$, which places it clearly in the IMBH regime (Fig.~\ref{XRFP}).  This, then, presents a clear example that ULXs with unusual properties can be powered by a different class of underlying compact object.

\section{Conclusions}
In our recent work we have shown that the main behaviours of many nearby ULXs can be understood in the context of super-Eddington accretion, where the anisotropy induced by an inhomogeneous and optically-thick radiatively-driven wind is the key component in shaping what we see.  Direct evidence for a wind is now emerging, both in our work and, for example, in a recent optical study where the spectra of several ULXs are shown to bear remarkable similarities to hot wind-dominated optical spectra, such as that from SS 433 (Fabrika et al. 2015).  However, there are challenges ahead for this model, not least how one can create the positive rms-flux variability relation and soft lags in NGC 5408 X-1 reported by Hern{\'a}ndez-Garc{\'i}a et al. (2015) from variability induced by a clumpy wind.  Also, with the imminent launch of {\it Astro-H\/} we will have access to the first calorimeter spectra of ULXs, from which we should learn a great deal about the putative wind material and its dynamics and physical state.

We also know now that ULXs are a heterogeneous population.  So, by finding objects that do not behave in the set patterns we see from objects in the ultraluminous state we can start to identify these minority populations, and hence we can finally begin to reveal the IMBH population that has long been suspected to play some part in explaining ULXs.

\acknowledgements 
TPR acknowledges support from STFC as part of the consolidated grant ST/L00075X/1.  MM acknowledges financial support from NASA {\it Chandra\/} Grant G05-16099X.  All data used is, or will be, public in the relevant mission archives.




\end{document}